# EMPIRICAL SCALING LAWS OF ROCKET EXHAUST CRATERING

Carly M. Donahue

Department of Physics, Astronomy and Geology, Berry College 2277 Martha Berry Hwy. NW, Mount Berry, GA 30149, cdonahue@berry.edu

Philip T. Metzger

The KSC Applied Physics Laboratory, John F. Kennedy Space Center, NASA YA-C3-E, Kennedy Space Center, Florida 32899, Philip.T.Metzger@nasa.gov

Christopher D. Immer

ASRC Aerospace, EDL Rm. 170h M/S ASRC-10, Kennedy Space Center, Florida 32899, Christopher.Immer-1@ksc.nasa.gov

(Dated: September 20, 2005)

When launching or landing a space craft on the regolith of a terrestrial surface, special attention needs to be paid to the rocket exhaust cratering effects. If the effects are not controlled, the rocket cratering could damage the spacecraft or other surrounding hardware. The cratering effects of a rocket landing on a planet's surface are not understood well, especially for the lunar case with the plume expanding in vacuum. As a result, the blast effects cannot be estimated sufficiently using analytical theories. It is necessary to develop physics-based simulation tools in order to calculate mission-essential parameters. In this work we test out the scaling laws of the physics in regard to growth rate of the crater depth. This will provide the physical insight necessary to begin the physics-based modeling.

## I. INTRODUCTION

During the Apollo and Viking programs there was considerable research into the blast effects of launching and landing on planetary regoliths[1–10]. That work ensured the success of those programs but also demonstrated that cratering will be a significant challenge for other mission scenarios. The problem will be more severe on Mars than on the Moon for three reasons. First, the Martian surface gravity is greater, requiring a greater thrust and a lower altitude at engine cutoff. Second, the thin Martian atmosphere collimates the supersonic jet of exhaust gas so that it is more focused on the regolith than if the plume were expanding into the lunar vacuum. Third, the lunar soil is (in many places, at least) extremely compacted below the top several centimeters, so that it acts as a very stable landing surface. Martian soil is disturbed by the wind and volatile cycles so that it is not expected to be as compacted. Because Martian landings with human-scaled spacecraft are expected to produce cratering much more severe than during Apollo or Viking missions, it is necessary to understand the physics of the cratering phenomena so that we can safely control the effects.

The lunar case is not expected to be as severe. However, a better understanding of lunar cratering is needed, too, because the lateral spray of the uppermost portion of the soil (the top several centimeters) still poses a challenge when multiple spacecraft attempts to land within short distances of one another. The first spacecraft to land may be scoured and contaminated by the spray from the second. There is relevant experience from when the Apollo 12 Lunar Module (LM) landed 155 meters away from the deactivated Surveyor 3 spacecraft. Portions of the Surveyor were then returned to Earth by the Apollo astronauts for analysis. It was found that the surfaces had been sandblasted and pitted and that every opening and joint was injected with grit from the high-speed spray[11,12].

There has been some important work scaling the erosion processes when a jet impinges a granular surface[13,14].

However, some key aspects of the physical scaling have not been determined, including the dependence of cratering rate upon the jet's gas density or upon other physical parameters. This project therefore began with a series of tests to simply observe and describe the physics and obtain these scaling laws. Future work will use these physical insights to begin computer modeling of the physics and adapting the model to lunar conditions to evaluate mitigation technologies.

Rajaratnam and Beltaos[14] introduced the erosion parameter $E_c$, based on the densimetric Froude number $Fr$, to predict the scaling of the final size of erosion craters. The Froude number and erosion parameter are defined in terms of the speed of the gas $V$, the density of the gas $\rho_g$, the density of the sand $\rho_s$, the diameter of the sand grains $d$, the height of the nozzle above the surface $H$, and the diameter of the nozzle $D$. These parameters are related by

$$Fr = \frac{\rho_g V^2}{(\rho_s - \rho_g)d}$$

$$E_c = \sqrt{Fr} \times \frac{D}{H}$$

In this work we test the erosion number against the velocity of the gas and the density of the gas to determine how well the erosion number can scale the growth rate of a crater formed by rocket exhaust from vertically impinging jets.

## II. SETUP

Our craters are formed in a box of sand with a metal pipe for the nozzle exiting a few inches above the sand. In order to be able to measure the depth of a crater, the nozzle is positioned directly above the sharp edge of a Plexiglas sheet. Therefore only the back half of the crater is formed behind the Plexiglas, allowing us to see the cratering process beneath the surface. The other sides of the box are made of metal. The box has dimensions of 12 inches by 12 inches by 7 inches. The Plexiglas extends higher than the other sides of the box in order to prevent grains from being ejected over the front, which would otherwise obscure the view. The Plexiglas has a cutout in it to allow for the nozzle to fit with little obstruction to air flow. The edge of the Plexiglas right below the nozzle is shaped to have a sharp edge in order to split the flow as smoothly as possible. We are assuming that the crater formed against the Plexiglas would be a true cross-section of a crater as it would have formed without the Plexiglas; however, the airflow over the Plexiglas has not been modeled. The nozzle extends vertically above the crater for about four feet so as to minimize obstructions such as bends right before the nozzle exit. Various gases are used (Helium, Nitrogen, Argon, and Carbon Dioxide) with carefully controlled mass flow rates. The height of the nozzle above the gas is three inches and the diameter of the nozzle is 3/8 inches. The box is filled with a quartz sand containing a very small fraction of dark grains, presumably muscovite. The particle sizes of the sand range between 180 and 200 microns. The specific gravity is measured at $2.6 \pm 0.1 \frac{kg}{liter}$ in agreement with the density of quartz. The density of the sand is measured to be $1.41 \pm 0.1$ $\frac{kg}{liter}$, therefore the packing fraction is 54%. The angle at failure is $35.5 \pm 1.5°$ and the angle of repose is $1.9 \pm 0.8°$.

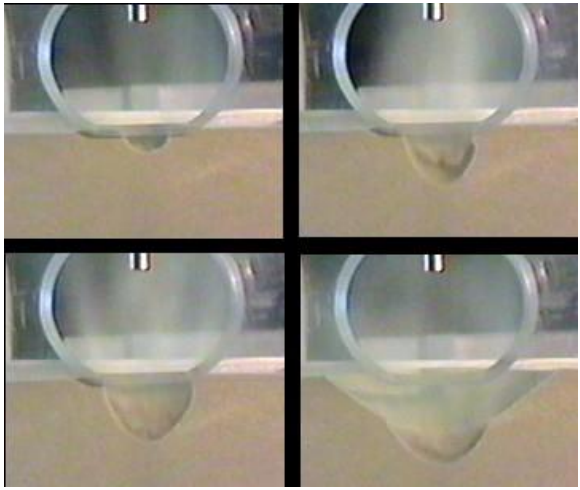

FIG. 1: The inner crater forms first until the sides are too steep and the crater is too deep. Then the sides fall in to create the outer crater.

A high speed camera is set up infront of the Plexiglass and films the entire cratering process. Using a machine vision software algorithms, data about the crater depth, width and volume can be extracted. The software can extract the data from every frame of film and at a frame speed of one hundred frames per second.

When the crater forms, two different shapes appear as shown in Figure 1. First, there is the inner crater in the shape of a parabola. The nozzle pressure is directly excavating the sand in the inner crater and ejecting it out. The shape of the inner crater causes a large part of the ejected sand to be directed straight up. Many parts along the sides of the inner crater are steeper than the angle of repose of the sand. The traction from the airflow is sufficient at those points to keep it at a steep angle. However, higher up the inner crater, the angle is too steep 2 and the traction from airflow is not sufficient enough to keep the sand from falling. At this point the outer crater forms. The sides of the outer crater are predicted to be the angle of repose of the sand. On the surface of the outer crater the sand rolls down to the center and into the airflow to be ejected up. Part of the sand ejected from the jet in the inner crater lands back into the outer crater to be ejected again.

## III. <u>FUNCTIONAL DEPENDENCE</u>

The scaling law for crater depth is simple despite the complex phenomenology. For instance, on the higher thrust tests, a deep inner crater forms first quickly and the the outer crater slowly forms. On the lower thrust tests, there is no initial large inner crater; instead, there is just a slow formation of both the inner crater and the outer crater at once. Despite these differences, the depth of the crater always grows as a logarithm of time over the full duration of the test, regardless of the gas velocity, height of the nozzle exit plane, and/or density of the gas. The large scale logarithmic behavior is perturbed by periodic avalanches on the crater's outer slope. While Rajarantnam and Beltaos have already noticed how crater depth and crater width both grow linearly with the logarithm of time, we additionally scale the growth parameters in terms of physical constants. The function that models the cratering has only two parameters.

$$d/a = \log\left(b\left(t + \frac{1}{b}\right)\right), t \geq 0$$

where $d$ is depth, and $t$ is time. The parameter $a$ scales the depth and $b$ scales the time. Here, the time shift is always $-\frac{1}{b}$. This equation implies that the excavation process is dominated by the simple differential equation

$$\dot{d}/a = be^{-d/a}$$

In Figure 2, graphs of the crater depth as a function of the logarithm of time for various gas densities and speeds are shown. Each graph represents one type of gas. Within each graph, the green function represents the gas at a speed of 56 m/s, the blue at 50 m/s, the purple at 40 m/s, and the red at 37 m/s. (For images in black and white, the green function is always on top, the blue is second from top, the purple is third from top, and the red is the bottom function.)

It is evident how each of the plots fits strikingly well to a logarithmic curve. For each of the plots, $a$ does not significantly vary. Therefore it can be assumed that $a$ is not dependent upon the speed of the gas or the density of the gas. However, the parameter $b$ does change significantly for both the speed of the gas and the density of the gas. In Figure 3, $a$ and $b$ are plotted against $\rho v^2$. Here, $a$ is constant with respect to $\rho v^2$, but $b$ has a positive linear relationship. The linear regression fit gives $a = 0.458$ in and $b = \left(0.00112 \frac{l \cdot s}{g \cdot m}\right) \rho v^2$.

We are currently in the process of testing the other parameters that the erosion number is depedent upon. Once all the testing is completed, then the Froude number and the erosion number can be evaluated as to their ability to predict the growth of rocket cratering on a terrestrial surface. The results indicate that the erosion number may not accurately characterize the crater formation. Currently we are in the process of measuring the growth of the crater with varying grain sizes and densities, and nozzle heights and depths. In addition to 3
the testing the other parameters, we will measure the sand displaced out of the crater and the crater width over time.

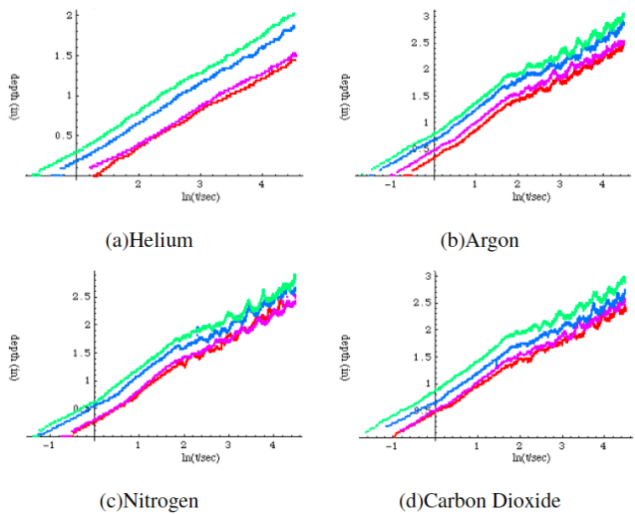

FIG. 2: The depth of the crater over time using various gases. Each plot has been time shifted by the fitting parameter $\frac{1}{b}$.

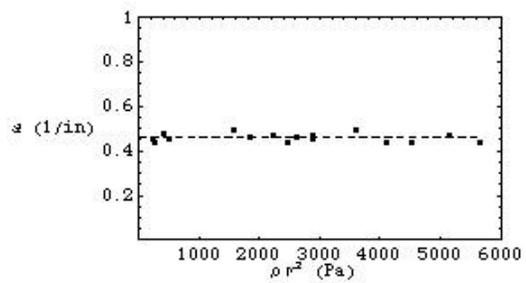

(a) $a$ versus $\rho v^2$

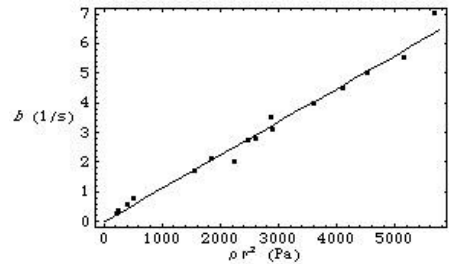

(b) $b$ versus $\rho v^2$

FIG. 3: Fitting growth parameters versus $\rho v^2$